\documentclass[
    ,final            % use final for the camera ready runs
    ,sort&compress    % use to sort references in citations
    ,numberedheadings % for numbered sections
  ]
  {aipproc}
\usepackage{hyperref}           % produces hyperlinked pdf documents
\usepackage[latin1]{inputenc}   % only if you use accented, non-english, characters
\layoutstyle{8x11single} \providecommand{\email}[1]{E-mail:
\href{mailto:#1}{\textnormal{\texttt{#1}}}}

\usepackage{epsf}
\usepackage{graphicx}

\begin{document}

\title{Low-Energy Quantum Gravity Leads \\ to Another
Picture of the Universe} 
\author{Michael A. Ivanov}{
  address={Physics Dept.,\\
Belarus State University of Informatics and Radioelectronics, \\
6 P. Brovka Street,  BY 220027, Minsk, Republic of Belarus.\\
\email{ivanovma@gw.bsuir.unibel.by}}}

\begin{abstract}
If gravitons are super-strong interacting particles and the
low-temperature graviton background exists, the basic cosmological
conjecture about the Dopplerian nature of redshifts may be false:
a full magnitude of cosmological redshift would be caused by
interactions of photons with gravitons. Non-forehead collisions
with gravitons will lead to a very specific additional relaxation
of any photonic flux that gives a possibility of another
interpretation of supernovae 1a data - without any kinematics.
These facts may implicate a necessity to change the standard
cosmological paradigm. Some features of a new paradigm are
discussed. In a frame of this model, every observer has two
different cosmological horizons. One of them is defined by maximum
existing temperatures of remote sources - by big enough distances,
all of them will be masked with the CMB radiation. Another, and
much smaller, one depends on their maximum luminosity - the
luminosity distance increases with a redshift much quickly than
the geometrical one.
\par If the considered quantum mechanism of classical gravity is realized
in the nature, then an existence of black holes contradicts to the
equivalence principle. In this approach, the two fundamental
constants - Hubble's and Newton's ones - should be connected
between themselves. The theoretical value of the Hubble constant
is computed. Also, every massive body would be decelerated due to
collisions with gravitons that may be connected with the Pioneer
10 anomaly.
\end{abstract}
\maketitle
\section[1]{Introduction }
An opinion is commonly accepted that quantum gravity should
manifest itself only on the Planck scale of energies, i.e. it is a
high-energy phenomenon. The value of the Planck energy $\sim
10^{19}$ GeV has been got from dimensional reasonings. Still one
wide-spread opinion is that we know a mechanism of gravity (bodies
are exchanging with gravitons of spin 2) but cannot correctly
describe it.

\par In a few last years, the situation has been abruptly changed. I
enumerate those discoveries and observations which may force, in
my opinion, the ice to break up. \par 1. In 1998, Anderson's team
reported about the discovery of anomalous acceleration of NASA's
probes Pioneer 10/11 \cite{1}; this effect is not embedded in a
frame of the general relativity, and its magnitude is somehow
equal to $\sim Hc$, where $H$ is the Hubble constant, $c$ is the
light velocity. \par 2. In the same 1998, two teams of
astrophysicists, which were collecting supernovae 1a data with the
aim to specificate parameters of cosmological expansion, reported
about dimming remote supernovae \cite{2,3}; the one would be
explained on a basis of the Doppler effect if at present epoch the
universe expands with acceleration. This explanation needs an
introduction of some "dark energy" which is unknown from any
laboratory experiments. \par 3. In January 2002, Nesvizhevsky's
team  reported about discovery of quantum states of ultra-cold
neutrons in the Earth's gravitational field \cite{4}. Observed
energies of levels (it means that and their differences too) in
full agreement with quantum-mechanical calculations turned out to
be equal to $\sim 10^{-12}$ eV. The formula for energy levels had
been found still by Bohr and Sommerfeld. If transitions between
these levels are accompanied with irradiation of gravitons then
energies of irradiated gravitons should have the same order - but
it is of 40 orders lesser than the Planck energy by which one
waits quantum manifestations of gravity. \par The first of these
discoveries obliges to muse about the borders of applicability of
the general relativity, the third - about that quantum gravity
would be a high-energy phenomenon. It seems that the second
discovery is far from quantum gravity but it obliges us to look at
the traditional interpretation of the nature of cosmological
redshift critically. An introduction into consideration of an
alternative model of redshifts \cite{5} which is based on a
conjecture about an existence of the graviton background gives us
odds to see on the effect of supernova dimming  as an additional
manifestation of low-energy quantum gravity. Under the definite
conditions, an effective temperature of the background may be the
same one as a temperature of the cosmic microwave background, with
an average graviton energy of the order of $\sim 10^{-3}$ eV. \par
In this contribution (it is a short version of my summarizing
paper \cite{500}), the main results of author's research in this
approach are described. It is shown that if a redshift would be a
quantum gravitational effect then one can get from its magnitude
an estimate of a new dimensional constant characterizing a single
act of interaction in this model. It is possible to calculate
theoretically a dependence of a light flux relaxation on a
redshift value, and this dependence fits supernova observational
data very well at least for $z < 0.5$. Further, it is possible to
find a pressure of single gravitons of the background which acts
on any pair of bodies due to screening the graviton background
with the bodies \cite{6}. It turns out that the pressure is huge
(a corresponding force is $\sim 1000$ times stronger than the
Newtonian attraction) but it is compensated with a pressure of
gravitons which are re-scattered by the bodies. The Newtonian
attraction arises if a part of gravitons of the background forms
pairs which are destructed by interaction with bodies. It is
interesting that both Hubble's and Newton's constants may be
computed in this approach with the ones being connected between
themselves. It allows us to get a theoretical estimate of the
Hubble constant. An unexpected feature of this mechanism of
gravity is a necessity of "an atomic structure" of matter - the
mechanism doesn't work without the one.
\par Collisions with gravitons should also call forth a deceleration
of massive bodies of order $\sim Hc$ - namely the same as of
NASA's probes. But at present stage it turns out unclear why such
the deceleration has {\it not} been observed for planets. The
situation reminds by something of the one that took place in
physics before the creation of quantum mechanics when a motion of
electrons should, as it seemed by canons of classical physics,
lead to their fall to a nucleus.
\par So, in this approach we deal with the following small quantum
effects of low-energy gravity: redshifts, its analog - a
deceleration of massive bodies, and an additional relaxation of
any light flux. The Newtonian attraction turns out to be the main
statistical effect, with bodies themselves being not sources of
gravitons - only correlational properties of {\it in} and {\it
out} fluxes of gravitons in their neighbourhood are changed due to
an interaction with bodies. There does still not exist a full and
closed theory in this approach, but even the initial researches in
this direction show that in this case quantum gravity cannot be
described separately of other interactions, and also manifest the
boundaries of applicability of a geometrical language in gravity.

\section[2]{  Passing photons through the graviton background \cite{5}}

Let us introduce the hypothesis, which is considered in this
approach as independent from the standard cosmological model:
there exists the isotropic graviton background. Photon scattering
is possible on gravitons $\gamma + h \to \gamma + h,$ where
$\gamma $ is a photon and $h$ is a graviton, if one of the
gravitons is virtual. The energy-momentum conservation law
prohibits energy transfer to free gravitons. Due to forehead
collisions with gravitons, an energy of any photon should decrease
when it passes through the sea of gravitons.
\par
From another side, none-forehead collisions of photons with
gravitons of the background will lead to an additional relaxation
of a photon flux, caused by transmission of a momentum transversal
component to some photons. It will lead to an additional dimming
of any remote objects, and may be connected with supernova
dimming.
\par We deal here with the
uniform non-expanding universe with the Euclidean space, and there
are not any cosmological kinematic effects in this model.
\subsection[2.1]{  Forehead collisions with gravitons: an alternative
explanations of the redshift nature} We shall take into account
that a gravitational "charge" of a photon must be proportional to
$E$ (it gives the factor  $E^{2}$ in a cross-section) and a
normalization of a photon wave function gives the factor $E^{-1}$
in the cross-section. Also we assume here that a photon average
energy loss $\bar \epsilon $ in one act of interaction is
relatively small to a photon energy $E.$ Then average energy
losses of a photon with an energy  $E $ on a way $dr $ will be
equal to \cite{5}:
\begin{equation}
                  dE=-aE dr,
\end{equation}
where $a$ is a constant. If a {\it whole} redshift magnitude is
caused by this effect, we must identify $a=H/c,$ where $c$ is the
light velocity, to have the Hubble law for small distances
\cite{104}.
\par
A photon energy $E$ should depend on a distance from a source $r$
as
\begin{equation}
                      E(r)=E_{0} \exp(-ar),
\end{equation}
where $E_{0}$ is an initial value of energy.
\par
The expression (2) is just only so far as the condition $\bar
\epsilon << E(r)$ takes place. Photons with a very small energy
may lose or acquire an energy changing their direction of
propagation after scattering.  Early or late such photons should
turn out in the thermodynamic equilibrium with the graviton
background, flowing into their own background. Decay of virtual
gravitons should give photon pairs for this background, too.
Perhaps, the last one is the cosmic microwave background \cite
{133,134}.
\par
It follows from the expression (2) that an exact dependence $r(z)$
is the following one:
\begin{equation}
                     r(z)= ln (1+z)/a,
\end{equation}
if an interaction with the graviton background is the only cause
of redshifts. It is very important, that this redshift does not
depend on a light frequency. For small $z,$ the dependence $r(z)$
will be linear.
\par
The expressions (1) - (3) are the same that appear in other
tired-light models (compare with \cite {212}). In this approach,
the ones follow from a possible existence of the isotropic
graviton background, from quantum electrodynamics, and from the
fact that a gravitational "charge" of a photon must be
proportional to $E.$
\par
There are two difficulties in this model: (i) possible blurring of
point sources, and (ii) the rise to fall time of the light curve
for SN1a sources (the relativistic time dilation effect)
\cite{2066} is in apparent conflict with the ideas here and of any
"tired light" explanation of effect. The problems are now open. I
think  that for (i), a solution will be based on the following: an
average graviton energy is not equal to zero, and after multiple
non-forehead collisions, photons should be rejected from the
registered flux - without a loss of definition of the source. For
(ii), a discrete character of photon energy losses (for some more
details, see the very end of subsection 2.3) would lead to a
deformation of the light curve and may give something like to time
dilation - but now, it is only the idea.

\subsection[2.2]{  Non-forehead collisions with gravitons: an
additional dimming of any light flux} Photon flux's average energy
losses on a way $dr$ due to non-forehead collisions with gravitons
should be proportional to $badr,$ where $b$ is a new constant of
the order $1.$ These losses are connected with a rejection of a
part of photons from a source-observer direction. Such the
relaxation together with the redshift will give a connection
between visible object's diameter and its luminosity (i.e. the
ratio of an  object visible angular diameter to a square root of
visible luminosity), distinguishing from the one of the standard
cosmological model.
\par
Let us consider that in a case of a non-forehead collision of a
graviton with a photon, the latter leaves a photon flux detected
by a remote observer (an assumption of a narrow beam of rays). The
details of calculation of the theoretical value of relaxation
factor $b$ which was used in author's paper \cite{5} were given
later in the preprint \cite{213}. So as both particles have
velocities $c,$ a cross-section of interaction, which is "visible"
under an angle $\theta$ (see Fig. 1), will be equal to $\sigma_{0}
\vert \cos \theta \vert$ if $\sigma_{0}$ is a cross-section by
forehead collisions. The function $\vert \cos \theta \vert$ allows
to take into account both front and back hemispheres for riding
gravitons. Additionally, a graviton flux, which falls on a picked
out area (cross-section), depends on the angle $\theta.$ We have
for the ratio of fluxes:
$$\Phi(\theta)/\Phi_{0}=S_{s}/\sigma_{0}, $$
where $\Phi(\theta)$ and $\Phi_{0}$ are the fluxes which fall on
$\sigma_{0}$ under the angle $\theta$ and normally, $S_{s}$ is a
square of side surface of a truncated cone with a base
$\sigma_{0}$ (see Fig. 1).

\begin{figure}[th]
\epsfxsize=12.98cm \centerline{\epsfbox{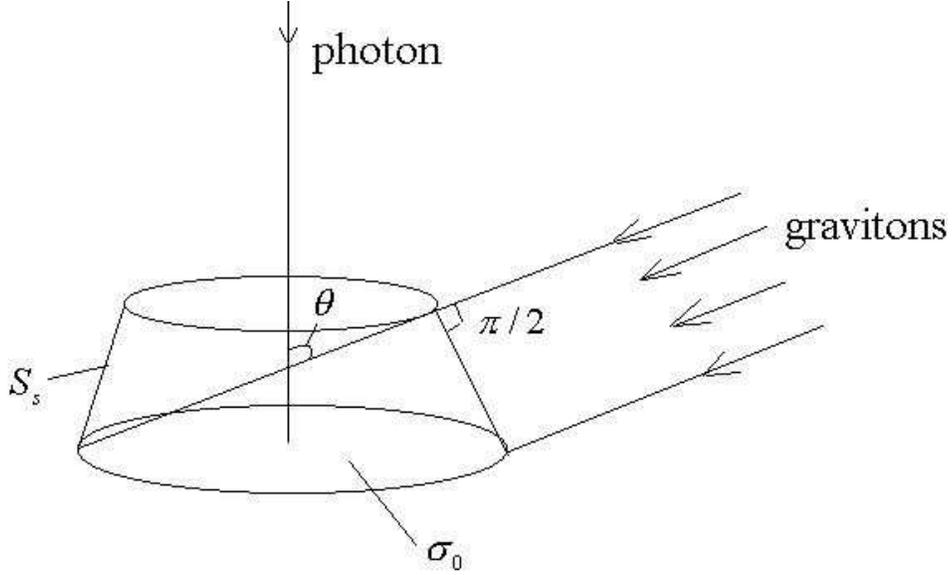}} \caption{By
non-forehead collisions of gravitons with a photon, it is
necessary to calculate a cone's side surface square, $S_{s}.$}
\end{figure}
Finally, we get for the factor $b:$
\begin{equation}
b=2 \int_{0}^{\pi/2}\cos{\theta}\times (S_{s}/\sigma_{0})\frac
{d\theta}{\pi/2}.
\end{equation}
By $0<\theta<\pi/4,$ a formed cone contains self-intersections,
and it is $S_{s}=2\sigma_{0} \times \cos{\theta}$. By
$\pi/4\leq\theta\leq\pi/2,$ we have $S_{s}=4\sigma_{0} \times
\sin^{2}{\theta}\cos{\theta}$.
\par
After computation of simple integrals, we get:
\begin{equation}
b=\frac {4}{\pi} (\int_{0}^{\pi/4}2\cos^{2}{\theta}d\theta +
\int_{\pi/4}^{\pi/2}\sin^{2}{2\theta}d\theta)= \frac {3}{2} +
\frac {2}{\pi} \simeq 2.137.
\end{equation}
In the considered simplest case of the uniform non-expanding
universe with the Euclidean space, we shall have the quantity
$$(1+z)^{(1+b)/2} \equiv (1+z)^{1.57}$$
in a visible object diameter-luminosity connection if a whole
redshift magnitude would caused by such an interaction with the
background (instead of $(1+z)^{2}$ for the expanding uniform
universe). For near sources, the estimate of the factor $b$ will
be some increased one.
\par
The luminosity distance (see \cite{2}) is a convenient quantity
for astrophysical observations. Both redshifts and the additional
relaxation of any photonic flux due to non-forehead collisions of
gravitons with photons lead in our model to the following
luminosity distance $D_{L}:$
\begin{equation}
D_{L}=a^{-1} \ln(1+z)\times (1+z)^{(1+b)/2} \equiv a^{-1}f_{1}(z),
\end{equation}
where $f_{1}(z)\equiv \ln(1+z)\times (1+z)^{(1+b)/2}$.
\subsection[2.3]{  Comparison of the theoretical predictions with
supernova data }
 To compare a form of this predicted dependence
$D_{L}(z)$ by unknown, but constant $H$, with the latest
observational supernova data by Riess et al. \cite{203}, one can
introduce distance moduli $\mu_{0} = 5 \log D_{L} + 25 = 5 \log
f_{1} + c_{1}$, where $c_{1}$ is an unknown constant (it is a
single free parameter to fit the data); $f_{1}$ is the luminosity
distance in units of $c/H$.
\begin{figure}[th]
\epsfxsize=12.98cm \centerline{\epsfbox{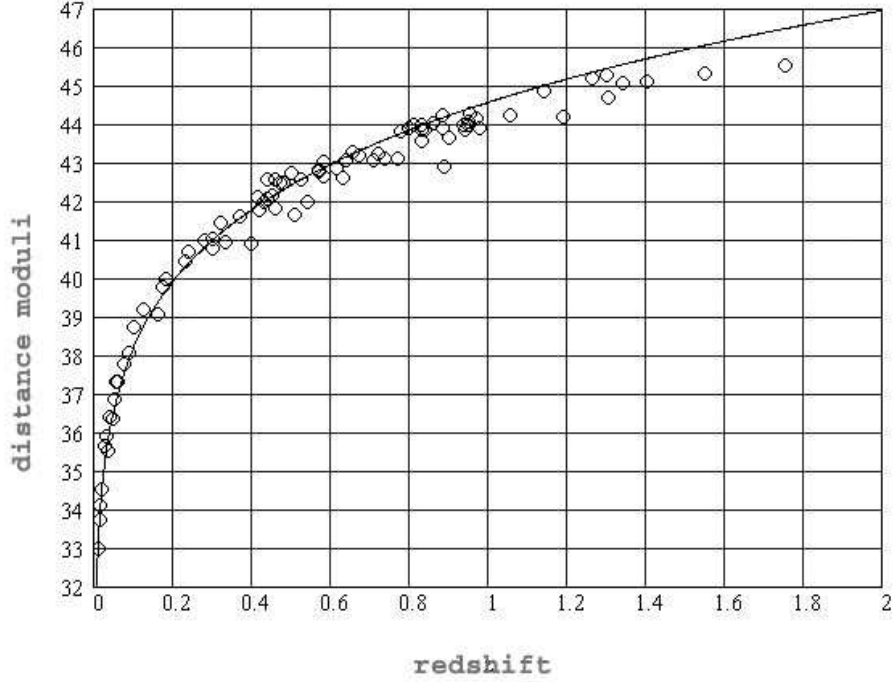}}
\caption{Comparison of the theoretical values of distance moduli
$\mu_{0}(z)$ in author's model (solid line) with observations
(points) from Table 5 of \cite{203} by Riess et al.}
\end{figure}
In Figure 2, the Hubble diagram $\mu_{0}(z)$ is shown with
$c_{1}=43$ to fit observations for low redshifts; observational
data (82 points) are taken from Table 5 of \cite{203}. The
predictions fit observations very well for roughly $z < 0.5$. It
excludes a need of any dark energy to explain supernovae dimming.
\par
Discrepancies between predicted and observed values of
$\mu_{0}(z)$ are obvious for higher $z$: we see that observations
show brighter SNe that the theory allows, and a difference
increases with $z$. It is better seen on Figure 3 with a linear
scale for $f_{1}$; observations are transformed as $\mu_{0}
\rightarrow 10^{(\mu_{0}-c_{1})/5}$ with the same
$c_{1}=43$.\footnote{A spread of observations raises with $z$; it
might be partially caused by quickly raising contribution of a
dispersion of measured flux: it should be proportional to
$f_{1}^{6}(z)$.}
\begin{figure}[th]
\epsfxsize=12.98cm \centerline{\epsfbox{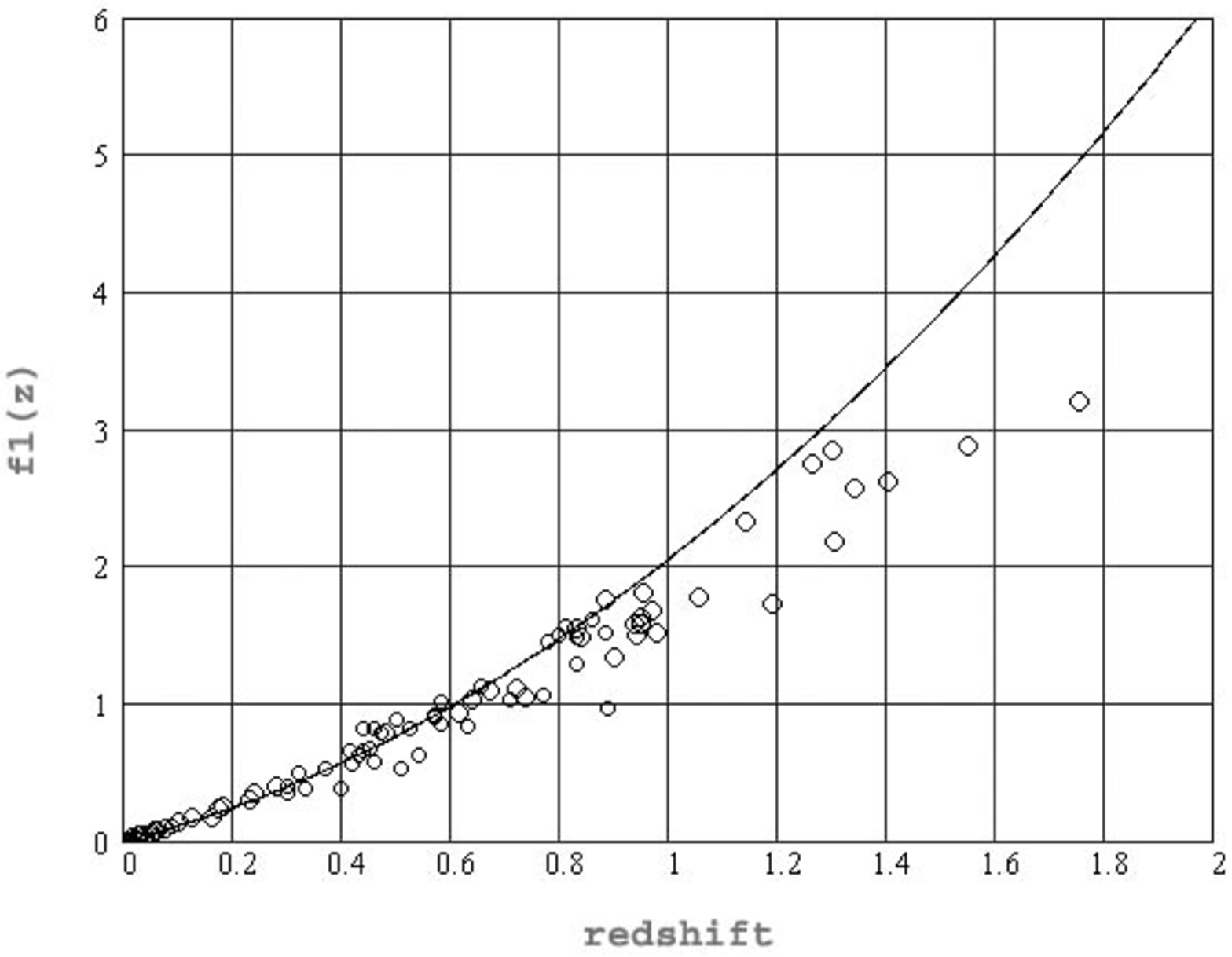}}
\caption{Predicted values of $f_{1}(z)$ (solid line) and
observations (points) from \cite{203} transformed to a linear
scale}
\end{figure}
It would be explained in the model as a result of specific
deformation of SN spectra due to a discrete character of photon
energy losses. Today, a theory of this effect does not exist, and
I explain its origin only qualitatively \cite{111}. For very small
redshifts $z,$ only a small part of photons transmits its energy
to the background (see Fig. 8 in \cite{500}). Therefore any
red-shifted narrow spectral strip will be a superposition of two
strips. One of them has a form which is identical to an initial
one, its space is proportional to $1-n(r)$ where $n(r)$ is an
average number of interactions of a single photon with the
background, and its center's shift is negligible (for a narrow
strip). Another part is expand, its space is proportional to
$n(r),$ and its center's shift is equal to $\bar \epsilon _{g} /h$
where $\bar \epsilon _{g} $   is an average energy loss in one act
of interaction. An amplitude of the red-shifted step should linear
raise with a redshift. For big $z,$ spectra of remote objects of
the universe would be deformed. A deformation would appear because
of multifold interactions of a initially-red-shifted part of
photons with the graviton background. It means that the observed
flux within a given passband would depend on a form of spectrum:
the flux may be larger than an expected one without this effect if
an initial flux within a next-blue neighbour band is big enough -
due to a superposition of red-shifted parts of spectrum. Some
other evidences of this effect would be an apparent variance of
the fine structure constant \cite{205} or of the CMB temperature
\cite{206} with epochs. In both cases, a ratio of red-shifted
spectral line's intensities may be sensitive to the effect. Also,
this effect should be taken into account when one analyzes a
temporal evolution of supernova spectra to detect the relativistic
"time dilation" effect \cite{2066}.

\subsection[2.4]{  Computation of the Hubble constant}
Let us consider that a full redshift magnitude is caused by an
interaction with single gravitons. If $\sigma (E,\epsilon)$ is a
cross-section of interaction by forehead collisions of a photon
with an energy $E$ with a graviton, having an energy $\epsilon,$
we consider really (see (1)), that
$${d \sigma (E,\epsilon) \over E d \Omega} = const(E),$$
where $d \Omega$ is a space angle element, and the function
$const(x)$ has a constant value for any $x$. If $f(\omega,T)d
\Omega /2 \pi$ is a spectral density of graviton flux in the
limits of $d \Omega$ in some direction ($\omega$ is a graviton
frequency, $\epsilon =\hbar\omega$), i.e. an intensity of a
graviton flux is equal to the integral $ (d \Omega/2 \pi)
\int_{0}^{\infty} f(\omega,T)d \omega,$ $T$ is an equivalent
temperature of the graviton background, we can write for the
Hubble constant $H=ac,$ introduced in the expression (1):
$$H={1 \over 2\pi} \int_{0}^{\infty} \frac {\sigma (E,\epsilon)}{E}
f(\omega,T)d \omega.$$ If $f(\omega,T)$ can be described by the
Planck formula for equilibrium radiation, then $$
\int_{0}^{\infty} f(\omega,T)d \omega = \sigma T^{4},$$ where
$\sigma$ is the Stephan- Boltzmann constant. As carriers of a
gravitational "charge" (without consideration of spin properties),
gravitons should be described in the same manner as photons , i.e.
one can write for them:
$${d \sigma (E,\epsilon) \over \epsilon d\Omega} = const(\epsilon).$$
Now let us introduce a new dimensional constant $D$, so that for
forehead collisions:
\begin{equation}
\sigma (E,\epsilon)= D \times E \times \epsilon.
\end{equation}
Then
\begin{equation}
H= {1 \over 2\pi} D \times \bar \epsilon \times (\sigma T^{4}),
\end{equation}
where $\bar \epsilon$ is an average graviton energy. Assuming $T
\sim 3 {\mathrm K},~ \bar \epsilon \sim 10^{-4}~ {\mathrm eV},$
and $H = 1.6 \times 10^{-18}~ {\mathrm s^{-1}},$ we get the
following rough estimate for $D:$
$$D \sim 10^{-27}~{\mathrm m^{2}/eV^{2}},$$ (see below
Subsection \ref{subsec:4.3}
for more exact estimate of $D$ and for a theoretical estimate of
$H$) that gives us the phenomenological estimate of cross-section
by the same and equal $E$ and $\bar \epsilon$:
$$\sigma (E,\bar \epsilon) \sim 10^{-35}~{\mathrm m^{2}}.$$

\section[3]{  Deceleration of massive bodies: an analog of redshifts}

As it was reported by Anderson's team \cite{1} , NASA deep-space
probes (Pioneer 10/11, Galileo, and Ulysses) experience a small
additional constant acceleration, directed towards the Sun (the
Pioneer anomaly). Today, a possible origin of the effect is
unknown. It must be noted here that the reported direction of
additional acceleration may be a result of the simplest
conjecture, which was accepted by the authors to provide a good
fit for all probes. One should compare different conjectures to
choose the one giving the best fit.
\par
We consider here a deceleration of massive bodies, which would
give a similar deformation of cosmic probes' trajectories
\cite{5}. The one would be a result of interaction of a massive
body with the graviton background, but such an additional
acceleration will be directed against a body velocity.
\par
It follows from a universality of gravitational interaction, that
not only photons, but all other objects, moving relative to the
background, should lose their energy, too, due to such a quantum
interaction with gravitons. If $a=H/c,$ it turns out that massive
bodies must feel a constant  deceleration of the same order of
magnitude as a small additional acceleration of cosmic probes.
\par
Let us now denote as $E$ a full energy of a moving body which has
a velocity $v$ relative to the background. Then energy losses of
the body by an interaction with the graviton background (due to
forehead collisions with gravitons) on the way $dr$ must be
expressed by the same formula (1):
$$ dE=-aE dr,$$
where $a=H/c.$ If $dr=vdt,$ where $t$ is a time, and
$E=mc^{2}/\sqrt{1-v^{2}/c^{2}},$ then we get for the body
acceleration $w \equiv dv/dt$ by a non-zero velocity:
\begin{equation}
w = - ac^{2}(1-v^{2}/c^{2}).
\end{equation}
We assume here, that non-forehead collisions with gravitons give
only stochastic deviations of a  massive body's velocity
direction, which are negligible. For small velocities:
\begin{equation}
w \simeq - Hc.
\end{equation}
If the Hubble constant $H$ is equal to $2.14 \times
10^{-18}{\mathrm s^{-1}}$ (it is the theoretical estimate of $H$
in this approach, see below Subsection \ref{subsec:4.3}), a
modulus of the acceleration will be equal to
\begin{equation}
|w| \simeq Hc = 6.419 \times 10^{-10} \ {\mathrm m/s^{2}},
\end{equation}
that has the same order of magnitude as  a value of the observed
additional acceleration $(8.74 \pm 1.33) \times 10^{-10} {\mathrm
m/s^2}$ for NASA probes.
\par
I must emphasize here that the acceleration $w$ is directed
against a body velocity only in a special frame of reference (in
which the graviton background is isotropic). I would like to note
that a deep-space mission to test the discovered anomaly is
planned now at NASA by the authors of this very important
discovery \cite{317}.
\par
It is very important to understand, why such an acceleration has
not been observed for planets. This acceleration will have
different directions by motion of a body on a closed orbit, and
one must take into account a solar system motion, too. As a
result, an orbit should be deformed. The observed value of
anomalous acceleration of Pioneer 10 should represent the vector
difference of the two accelerations \cite{6}: an acceleration of
Pioneer 10 relative to the graviton background, and  an
acceleration of the Earth relative to the background. Possibly,
the last is displayed as an annual periodic term in the residuals
of Pioneer 10 \cite{311}. If the solar system moves with a
noticeable velocity relative to the background, the Earth's
anomalous acceleration projection on the direction of this
velocity will be smaller than for the Sun - because of the Earth's
orbital motion. It means that in a frame of reference, connected
with the Sun, the Earth should move with an anomalous acceleration
having non-zero projections as well on the orbital velocity
direction as on the direction of solar system motion relative to
the background. Under some conditions, the Earth's anomalous
acceleration in this frame of reference may be periodic. The axis
of Earth's orbit should feel an annual precession by it. This
question needs a further consideration.

\section[4]{  Gravity as the screening effect}
It was shown by the author \cite{6} that screening the background
of super-strong interacting gravitons creates for any pair of
bodies both attraction and repulsion forces due to pressure of
gravitons. For single gravitons, these forces are approximately
balanced, but each of them is much bigger than a force of
Newtonian attraction. If single gravitons are pairing, an
attraction force due to pressure of such graviton pairs is twice
exceeding a corresponding repulsion force if graviton pairs are
destructed by collisions with a body. In such the model, the
Newton constant is connected with the Hubble constant that gives a
possibility to obtain a theoretical estimate of the last. We deal
here with a flat non-expanding universe fulfilled with
super-strong interacting gravitons; it changes the meaning of the
Hubble constant which describes magnitudes of three small effects
of quantum gravity but not any expansion or an age of the
universe.
\subsection[4.1]{  Pressure force of single gravitons}
If gravitons of the background run against a pair of bodies with
masses $m_{1}$ and $m_{2}$ (and energies $E_{1}$ and $E_{2}$) from
infinity, then a part of gravitons is screened. Let $\sigma
(E_{1},\epsilon)$ is a cross-section of interaction of body $1$
with a graviton with an energy $\epsilon=\hbar \omega,$ where
$\omega$ is a graviton frequency, $\sigma (E_{2},\epsilon)$ is the
same cross-section for body $2.$ In absence of body $2,$ a whole
modulus of a gravitonic pressure force acting on body $1$ would be
equal to:
\begin{equation}
4\sigma (E_{1},<\epsilon>)\times {1 \over 3} \times {4 f(\omega,
T) \over c},
\end{equation}
where $f(\omega, T)$ is a graviton spectrum with a temperature $T$
(assuming to be Planckian), the factor $4$ in front of $\sigma
(E_{1},<\epsilon>)$ is introduced to allow all possible directions
of graviton running, $<\epsilon>$ is another average energy of
running gravitons with a frequency $\omega$ taking into account a
probability of that in a realization of flat wave a number of
gravitons may be equal to zero, and that not all of gravitons ride
at a body.
\par
Body $2,$ placed on a distance $r$ from body $1,$ will screen a
portion of running against body $1$ gravitons which is equal for
big distances between the bodies (i.e. by $\sigma
(E_{2},<\epsilon>) \ll 4 \pi r^{2}$) to:
\begin{equation}
\sigma (E_{2},<\epsilon>) \over 4 \pi r^{2}.
\end{equation}
Taking into account all frequencies $\omega,$ the following
attractive force will act between bodies $1$ and $2:$
\begin{equation}
F_{1}= \int_{0}^{\infty} {\sigma (E_{2},<\epsilon>) \over 4 \pi
r^{2}} \times 4 \sigma (E_{1},<\epsilon>)\times {1 \over 3} \times
{4 f(\omega, T) \over c} d\omega.
\end{equation}
Let $f(\omega, T)$ is described with the Planck formula:
\begin{equation}
f(\omega,T)={{\omega}^{2} \over {4{\pi}^{2} c^{2}}} {{\hbar
\omega} \over {\exp(\hbar \omega/kT) - 1}}.
\end{equation}
Let $x \equiv {\hbar \omega/  kT},$ and $\bar{n} \equiv {1/
(\exp(x)-1)}$ is an average number of gravitons in a flat wave
with a frequency $\omega$ (on one mode of two distinguishing with
a projection of particle spin). Let $P(n,x)$ is a probability of
that in a realization of flat wave a number of gravitons is equal
to $n,$ for example $P(0,x)=\exp(-\bar{n}).$
\par
Then we get for an attractive force $F_{1}:$
\begin{equation}
F_{1}= {4 \over 3}  {{D^{2} E_{1} E_{2}} \over {\pi r^{2} c}}
\int_{0}^{\infty} {{{\hbar}^{3} \omega^{5}} \over {4\pi^{2}c^{2}}}
[1-P(0,x)]^{2}\bar{n}^{5}\exp(-2\bar{n}) d\omega
\end{equation}
$$= {1 \over 3} \times {{D^{2} c (kT)^{6} m_{1} m_{2}} \over
{\pi^{3}\hbar^{3}r^{2}}} \times I_{1},$$ where
\begin{equation}
I_{1} \equiv \int_{0}^{\infty} x^{5}
(1-\exp\{-[\exp(x)-1]^{-1}\})^{2}[\exp(x)-1]^{-5}
\exp\{-2[\exp(x)-1]^{-1}\} dx
\end{equation}
$$= 5.636 \times 10^{-3}.$$ This and all other integrals were found
with the MathCad software.

If $F_{1}\equiv G_{1} \times  m_{1}m_{2}/r^{2},$ then the constant
$G_{1}$ is equal to:
\begin{equation}
G_{1} \equiv {1 \over 3} \times {D^{2} c(kT)^{6} \over
{\pi^{3}\hbar^{3}}} \times I_{1}.
\end{equation}
By $T=2.7~{\mathrm K}:$ $G_{1} =1215.4 \times G,$ that is three
order greater than the Newton constant, $G.$
\par
But if single gravitons are elastically scattered with body $1,$
then our reasoning may be reversed: the same portion (13) of
scattered gravitons will create a repulsive force $F_{1}^{'}$
acting on body $2$ and equal to $F_{1}^{'} =F_{1},$ if one
neglects with small allowances which are proportional to $D^{3}/
r^{4}.$
\par
So, for bodies which elastically scatter gravitons, screening a
flux of single gravitons does not ensure Newtonian attraction. But
for gravitonic black holes which absorb any particles and do not
re-emit them (by the meaning of a concept, the ones are usual
black holes; I introduce a redundant adjective only from a
caution), we will have $F_{1}^{'} =0.$ It means that such the
object would attract other bodies with a force which is
proportional to $G_{1}$ but not to $G,$ i.e. Einstein's
equivalence principle would be violated for them. This conclusion,
as we shall see below, stays in force for the case of graviton
pairing, too.

\subsection[4.2]{  Graviton pairing}
To ensure an attractive force which is not equal to a repulsive
one, particle correlations should differ for {\it in} and {\it
out} flux. For example, single gravitons of running flux may
associate in pairs \cite{6}. If such pairs are destructed by
collision with a body, then quantities $<\epsilon>$ will be
distinguished for running and scattered particles. Graviton
pairing may be caused with graviton's own gravitational attraction
or gravitonic spin-spin interaction. Left an analysis of the
nature of graviton pairing for the future; let us see that gives
such the pairing.
\par
To find an average number of pairs $\bar{n}_{2}$ in a wave with a
frequency $\omega$ for the state of thermodynamic equilibrium, one
may replace $\hbar \rightarrow 2\hbar$ by deducing the Planck
formula. Then an average number of pairs will be equal to:
\begin{equation}
\bar{n}_{2} ={1 \over {\exp(2x)-1}},
\end{equation}
and an energy of one pair will be equal to $2\hbar \omega.$ It is
important that graviton pairing does not change a number of
stationary waves, so as pairs nucleate from existing gravitons.
\par
It follows from the energy conservation law that composite
gravitons should be distributed only in two modes. So as
\begin{equation}
\lim_{x \to 0} {\bar{n}_{2} \over \bar{n}} ={1/2},
\end{equation}
then by $x \rightarrow 0$ we have $2\bar{n}_{2}=\bar{n},$ i.e. all
of gravitons are pairing by low frequencies. An average energy on
every mode of pairing gravitons is equal to $2 \hbar \omega
\bar{n}_{2},$ the one on every mode of single gravitons - to
$\hbar \omega \bar{n}.$ These energies are equal by $x \rightarrow
0,$ because of that, the numbers of modes are equal, too, if the
background is in the thermodynamic equilibrium with surrounding
bodies.
\par
The spectrum of composite gravitons is also the Planckian one, but
with a smaller temperature of $0.5946 T.$ An absolute luminosity
for the sub-system of composite gravitons is equal to ${1 \over
8}\sigma T^{4},$ where $\sigma$ is the Stephan-Boltzmann constant.
It is important that the graviton pairing effect does not change
computed values of the Hubble constant and of anomalous
deceleration of massive bodies: twice decreasing of a sub-system
particle number due to the pairing effect is compensated with
twice increasing the cross-section of interaction of a photon or
any body with such the composite gravitons. Non-pairing gravitons
with spin $1$ give also its contribution in values of redshifts,
an additional relaxation of light intensity due to non-forehead
collisions with gravitons, and  anomalous deceleration of massive
bodies moving relative to the background.

\subsection[4.3]{  Computation of the Newton constant, and a connection
between the two fundamental constants, $G$ and $H$ }
\label{subsec:4.3}  If running graviton pairs ensure for two
bodies an attractive force $F_{2},$ then a repulsive force due to
re-emission of gravitons of a pair alone will be equal to
$F_{2}^{'} =F_{2}/2.$ It follows from that the cross-section for
{\it single additional scattered} gravitons of destructed pairs
will be twice smaller than for pairs themselves (the leading
factor $2\hbar \omega$ for pairs should be replaced with $\hbar
\omega$ for single gravitons). For pairs, we introduce here the
cross-section $ \sigma (E_{2},<\epsilon_{2}>),$ where
$<\epsilon_{2}>$ is an average pair energy with taking into
account a probability of that in a realization of flat wave a
number of graviton pairs may be equal to zero, and that not all of
graviton pairs ride at a body ($<\epsilon_{2}>$ is an analog of
$<\epsilon>$). This equality is true in neglecting with small
allowances which are proportional to $D^{3}/ r^{4}$. Replacing
$\bar{n} \rightarrow \bar{n}_{2},~ \hbar \omega \rightarrow 2\hbar
\omega,$ and $P(n,x) \rightarrow P(n,2x),$ where $P(0,2x)=
\exp(-\bar{n}_{2}),$ we get for a force of attraction of two
bodies due to pressure of graviton pairs \footnote{In initial
version of this paper, factor 2 was lost in the right part of Eq.
(21), and the theoretical values of $D$ and $H$ were overestimated
of $\sqrt{2}$ times}, $F_{2}$:
\begin{equation}
F_{2}= \int_{0}^{\infty} {\sigma (E_{2},<\epsilon_{2}>) \over 4
\pi r^{2}} \times 4 \sigma (E_{1},<\epsilon_{2}>)\times {1 \over
3} \times {4 f_{2}(2\omega,T) \over c} d\omega
\end{equation}
$$= {8 \over 3} \times
{D^{2} c(kT)^{6} m_{1}m_{2} \over {\pi^{3}\hbar^{3}r^{2}}}\times
I_{2},$$ where
\begin{equation}
I_{2} \equiv \int_{0}^{\infty}{ x^{5}
(1-\exp\{-[\exp(2x)-1]^{-1}\})^{2}[\exp(2x)-1]^{-5} \over
\exp(2(\exp(2x)-1)^{-1}) \exp\{2[\exp(x)-1]^{-1}\}} d x
\end{equation}
$$= 2.3184 \times 10^{-6}.$$
The difference $F$ between attractive and repulsive forces will be
equal to:
\begin{equation}
F \equiv F_{2}- F_{2}^{'}={1 \over 2}F_{2} \equiv G_{2}{m_{1}m_{2}
\over r^{2}},
\end{equation}
where the constant $G_{2}$ is equal to:
\begin{equation}
G_{2} \equiv {4 \over 3} \times {D^{2} c(kT)^{6} \over
{\pi^{3}\hbar^{3}}} \times I_{2}.
\end{equation}
Both $G_{1}$ and $G_{2}$ are proportional to $T^{6}$ (and $H \sim
T^{5},$ so as $\bar{\epsilon} \sim T$).
\par
If one assumes that $G_{2}=G,$ then it follows that by $T=2.7
{\mathrm K}$ the constant $D$ should have the value:
\begin{equation}
D=0.795 \times 10^{-27}{\mathrm {m^{2} / eV^{2}}}.
\end{equation}
An average graviton energy of the background is equal to:
\begin{equation}
\bar{\epsilon} \equiv \int_{0}^{\infty} \hbar \omega \times
{f(\omega, T) \over \sigma T^{4}} d \omega = {15 \over
\pi^{4}}I_{4}kT,
\end{equation}
where
$$I_{4} \equiv \int_{0}^{\infty} {x^{4} dx \over
{\exp(x)-1}}=24.866 $$ (it is $\bar{\epsilon}=8.98 \times
10^{-4}{\mathrm eV}$ by $T=2.7 {\mathrm K}$).
\par
We can use (8) and (24) to establish a connection between the two
fundamental constants, $G$ and $H$, under the condition that
$G_{2}=G.$ We have for $D:$
\begin{equation}
D= {2\pi H \over \bar{\epsilon} \sigma T^{4}}= {2 \pi^{5} H \over
15 k \sigma T^{5} I_{4}};
\end{equation}
then
\begin{equation}
G=G_{2} = {4 \over 3} \times {D^{2} c(kT)^{6} \over
{\pi^{3}\hbar^{3}}} \times I_{2}= \\
{64 \pi^{5} \over 45} \times {H^{2}c^{3}I_{2} \over \sigma T^{4}
I_{4}^{2}}.
\end{equation}
So as the value of $G$ is known much better than the value of $H,$
let us express $H$ via $G:$
\begin{equation}
H= (G  {45 \over 64 \pi^{5}}  {\sigma T^{4} I_{4}^{2} \over
{c^{3}I_{2}}})^{1/2}= 2.14 \times 10^{-18}~{\mathrm s^{-1}},
\end{equation}
or in the units which are more familiar for many of us: $H=66.875
\ {\mathrm km \times s^{-1} \times Mpc^{-1}}.$
\par
This value of $H$ is in the good accordance with the majority of
present astrophysical estimations \cite{2,512,513} (for example,
the estimate $(72 \pm 8){\mathrm km/s/Mpc}$ has been got from SN1a
cosmological distance determinations in \cite{513}), but it is
lesser than some of them \cite{512a} and than it follows from the
observed value of anomalous acceleration of Pioneer 10 \cite{1}.

\section[5]{  Some cosmological consequences of the model}
If the described model of redshifts is true, what is a picture of
the universe? In a frame of this model, every observer has two own
spheres of observability in the universe (two different
cosmological horizons exist for any observer) \cite{44,55}. One of
them is defined by maximum existing temperatures of remote sources
- by big enough distances, all of them will be masked with the CMB
radiation. Another, and much smaller, sphere depends on their
maximum luminosity - the luminosity distance increases with a
redshift much quickly than the geometrical one. The ratio of the
luminosity distance to the geometrical one is the quickly
increasing function of $z:$
\begin{equation}
D_{L}(z)/r(z)= (1+z)^{(1+b)/2},
\end{equation}
which does not depend on the Hubble constant.  An outer part of
the universe will drown in a darkness. \par By the found
theoretical value of the Hubble constant: $H= 2.14 \times
10^{-18}~{\mathrm s^{-1}}$ (then a natural light unit of distances
is equal to $1/H \simeq 14.85$ light GYR), plots of two
theoretical functions of $z$ in this model - the geometrical
distance $r(z)$ and the luminosity distance $D_{L}(z)$ - are shown
on Fig. 4 \cite{44,55}. As one can see, for objects with $z \sim
10$, which are observable now, we should anticipate geometrical
distances of the order $\sim 35$ light GYR and luminosity
distances of the order $\sim 1555$ light GYR in a frame of this
model. An estimate of distances to objects with given $z$ is
changed, too: for example, the quasar with $z=5.8$ \cite {043}
should be in a distance approximately $2.8$ times bigger than the
one expected in the model based on the Doppler effect.
\begin{figure}[th]
\epsfxsize=12.98cm \centerline{\epsfbox{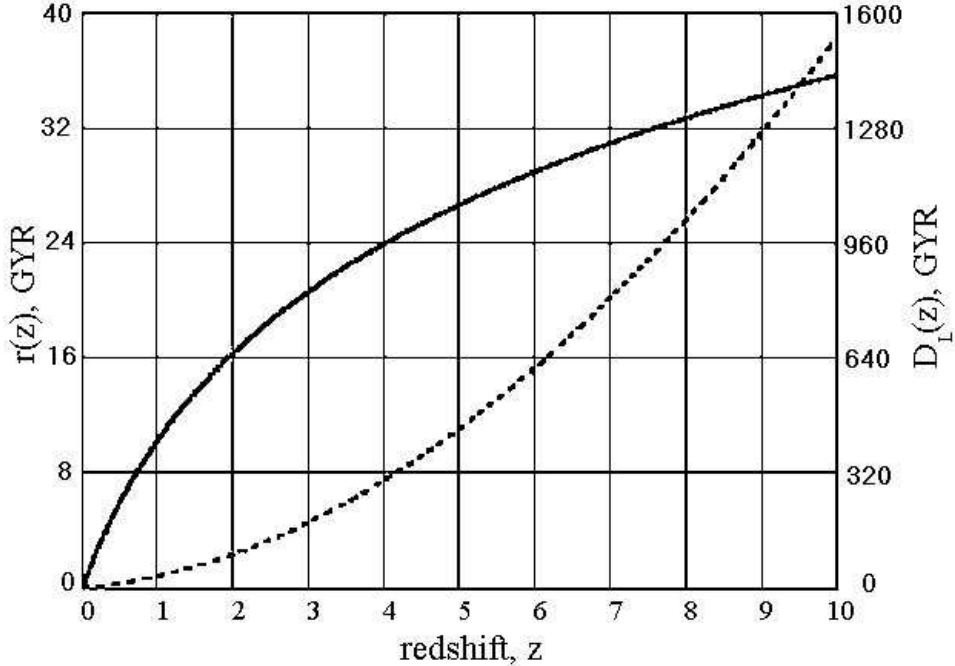}} \caption{The
geometrical distance, $r(z),$ (solid line) and the luminosity
distance, $D_{L}(z),$ (dashed line) - both in light GYRs - in this
model as functions of a redshift, z. The following theoretical
value for $H$ is accepted: $H= 2.14 \times 10^{-18} {\mathrm
s^{-1}}$.}
\end{figure}
\par
We can assume that the graviton background and the cosmic
microwave one are in a state of thermodynamical equilibrium, and
have the same temperatures. CMB itself may arise as a result of
cooling any light radiation up to reaching this equilibrium. Then
it needs $z \sim 1000$ to get through the very edge of our cosmic
"ecumene".
\par Some other possible cosmological
consequences of an existence of the graviton background were
described in \cite{08,6}. Observations of last years give us
strong evidences for supermassive and compact objects (named now
supermassive black holes) in active and normal galactic nuclei
\cite{65,67,68,612,616}. Massive nuclear "black holes" of $10^{6}
- 10^{9}$ solar masses may be responsible for the energy
production in quasars and active galaxies \cite{65}. In a frame of
this model, an existence of black holes contradicts to the
equivalence principle. It means that these objects should have
another nature; one must remember that we know only that these
objects are supermassive and compact.
\par
There should be two opposite processes of heating and cooling the
graviton background \cite{08} which may have a big impact on
cosmology. Unlike models of expanding universe, in any tired light
model one has a problem of utilization of energy, lost by
radiation of remote objects. In the considered model, a virtual
graviton forms under collision of a photon with a graviton of the
graviton background. It should be massive if an initial graviton
transfers its total momentum to a photon; it follows from the
energy conservation law that its energy $\epsilon^{'}$ must be
equal to $2 \epsilon$ if $\epsilon$ is an initial graviton energy.
In force of the uncertainty relation, one has for a virtual
graviton lifetime $\tau:$ $\tau \leq  \hbar/\epsilon^{'},$ i.e.
for $\epsilon^{'} \sim 10^{-4}~ {\mathrm eV}$ it is $\tau \leq
10^{-11}~{\mathrm s}.$ In force of conservation laws for energy,
momentum and angular momentum, a virtual graviton may decay into
no less than three real gravitons. In a case of decay into three
gravitons, its energies should be equal to $\epsilon,~
\epsilon^{''},~ \epsilon {'''},$ with $\epsilon^{''} + \epsilon
{'''}= \epsilon.$ So, after this decay, two new gravitons with
$\epsilon^{''},~ \epsilon {'''} < \epsilon$ inflow into the
graviton background. It is a source of adjunction of the graviton
background.
\par
From another side, an interaction of gravitons of the background
between themselves should lead to the formation of virtual massive
gravitons, too, with energies less than $\epsilon_{min}$ where
$\epsilon_{min}$ is a minimal energy of one graviton of an initial
interacting pair. If gravitons with energies $\epsilon^{''},~
\epsilon {'''}$ wear out a file of collisions with gravitons of
the background, its lifetime increases. In every such a
collision-decay cycle, an average energy of "redundant" gravitons
will double decrease, and its lifetime will double increase. Only
for $\sim 93$ cycles, a lifetime will increase from
$10^{-11}~{\mathrm s}$ to $10$ Gyr. Such virtual massive
gravitons, with a lifetime increasing from one collision to
another, would duly serve dark matter particles. Having a zero (or
near to zero) initial velocity relative to the graviton
background, the ones will not interact with matter in any manner
excepting usual gravitation. An ultra-cold gas of such gravitons
will condense under influence of gravitational attraction into
"black holes" or other massive objects. Additionally to it, even
in absence of initial heterogeneity, the one will easy arise in
such the gas that would lead to arising of super compact massive
objects, which will be able to turn out "germs" of "black holes".
It is a method "to cool" the graviton background.
\par
So, the graviton background may turn up "a perpetual engine" of
the universe, pumping energy from any radiation to massive
objects. An equilibrium state of the background will be ensured by
such a temperature $T,$ for which an energy profit of the
background due to an influx of energy from radiation will be equal
to a loss of its energy due to a catch of virtual massive
gravitons with "black holes" or other massive objects. In such the
picture, the chances are that "black holes" would turn out "germs"
of galaxies. After accumulation of a big enough energy by a "black
hole" (to be more exact, by a super-compact massive object) by
means of a catch of virtual massive gravitons, the one would be
absolved from an energy excess in via ejection of matter, from
which stars of galaxy should form. It awaits to understand else in
such the approach how usual matter particles form from virtual
massive gravitons. \par There is a very interesting but
non-researched possibility: due to relative decreasing of an
intensity of graviton pair flux in an internal area of galaxies
(pairs are destructed under collisions with matter particles), the
effective Newton constant may turn out to be running on galactic
scales. It might lead to something like to the modified Newtonian
dynamics (MOND) by  Mordehai Milgrom (about MOND, for example, see
\cite{99}). But to evaluate this effect, one should take into
account a relaxation process for pairs, about which we know
nothing today. It is obvious only that gravity should be stronger
on a galactic periphery.

\section[6]{  Conclusion}
It follows from the above consideration that the geometrical
description of gravity should be a good idealization for any pair
of bodies at a big distance by the condition of an "atomic
structure" of matter. This condition cannot be accepted only for
black holes which must interact with gravitons as aggregated
objects. In addition, the equivalence principle is roughly broken
for black holes, if the described quantum mechanism of classical
gravity is realized in the nature. Because attracting bodies are
not initial sources of gravitons, a future theory must be
non-local in this sense to describe gravitons running from
infinity. The Le Sage's idea to describe gravity as caused by
running {\it ab extra} particles was criticized by the great
physicist Richard Feynman in his public lectures at Cornell
University \cite{a15}, but the Pioneer 10 anomaly \cite{1},
perhaps, is a good contra argument pro this idea. \par The
described quantum mechanism of classical gravity is obviously
asymmetric relative to the time inversion. By the time inversion,
single gravitons would run against bodies to form pairs after
collisions with bodies. It would lead to replacing a body
attraction with a repulsion. But such the change will do
impossible the graviton pairing. \par A future theory dealing with
gravitons as usual particles should have a number of features
which are not characterizing any existing model to image the
considered here features of the possible quantum mechanism of
gravity. If this mechanism is realized in the nature, both the
general relativity and quantum mechanics should be modified. Any
divergencies, perhaps, would be not possible in such the model
because of natural smooth cut-offs of the graviton spectrum from
both sides. Gravity at short distances, which are much bigger than
the Planck length, needs to be described only in some unified
manner.

\end{document}